\documentclass{aa}
\usepackage{graphics}
\def\gtsima{$\; \buildrel > \over \sim \;$}
\begin{document}
\hyphenation{gra-ting pla-ne-ta-ry res-pec-ti-ve-ly mo-dels}

\thesaurus{05:10.15.1}

\title{Dissolving star cluster candidates}

\author{E. Bica \inst{1}, B.X. Santiago \inst{1}, C.M. Dutra \inst{1}, H. Dottori \inst{1}, M.R. de Oliveira \inst{1} \and D. Pavani \inst{1}}

\offprints{ C.M. Dutra - dutra@if.ufrgs.br}
 
\institute{Universidade Federal do Rio Grande do Sul, IF, 
CP\,15051, Porto Alegre 91501--970, RS, Brazil}

\date{Received 13.09.2000; accepted 14.11.2000}

\titlerunning{Dissolving star cluster candidates}
\authorrunning{E. Bica et al. }

\maketitle

\begin{abstract}

We present a list of  34 neglected entries from star cluster catalogues located at relatively high galactic latitudes ($|b| >$ 15$^{\circ}$) which appear to be candidate late stages of star cluster dynamical evolution.  Although underpopulated with respect to usual open clusters, they still present a high number density contrast as compared to the galactic field. This was verified by means  of (i) predicted model counts from different galactic subsystems in the same direction, and (ii) Guide Star Catalog equal solid angle counts for the object and surrounding fields.
This suggests that the  objects are physical systems, possibly  star clusters in the process of disruption or  their fossil remains.
The sample will be useful for followup studies in view of  verifying their physical nature.


\end{abstract}

\section{Introduction}

Star Clusters are known to dynamically evolve and stellar depletion effects eventually  lead to the cluster dissolution. Fundamental questions are: (i) where are the clusters in process of dissolution? (ii) if  fossils are left, can any remnant  be detected? The present study aims at showing that several candidates for these effects occur in star cluster catalogues themselves. 

Several poorly populated  objects  at relatively high galactic latitudes (from the  open cluster perspective $|b| >$ 15$^{\circ}$) are included in open cluster catalogues (Alter et al. 1970, Lyng\aa  \,1987). Other objects were reported as clusters  in  early studies (e.g. New General Catalogue and Index Catalogue), or in modern ones like the ESO (B) Atlas Survey Catalogue (Lauberts 1982). 

Depletion of Main Sequence (MS) stars  has been detected or evidence of it has been found in luminosity functions and Colour-Magnitude Diagrams (CMDs) of some Palomar or Palomar-like  globular clusters such as E3 (McClure et al 1985), ESO452-SC11
(Bica et al. 1999) and NGC6717 (Palomar 9) (Ortolani et al. 1999). Such evolved dynamical stages of low mass globular clusters are still associated to relatively well  populated star clusters, but one may wonder what subsequent stages would look like, probably an
underpopulated fossil core containing some double and multiple stars.

Among open clusters low  MS   depletion has been found e.g. in the intermediate age ($\approx$ 3 Gyr)  cluster NGC3680 (Anthony-Twarog et al. 1991).
The dissolution of open clusters has been studied by Wielen (1971). Updated data suggest that most open clusters dissolve in about 100 Myr and this will probably not change much as one includes fainter clusters  (Ahumada et al. 2000). Intermediate age open clusters are certainly survivors of initially massive clusters  (Friel 1995), and their  updated age histogram containing more than 100 entries (Dutra \& Bica 2000) suggests a dissolution timescale of about 1 Gyr. Also, N-body simulations of star clusters in an external potential have shown typical dissolution times in the range 500-2500 Myr (Terlevich 1987, McMillan \& Hut 1994, de la Fuente Marcos 1997, Portegies Zwart et al. 2000). Mass segregation is expected to occur in cluster cores during one relaxation time, according to N-body simulations (Terlevich 1987, Portegies Zwart et al. 2000). An important effect of mass segregation is the depletion of low mass MS stars by means of evaporation due to the tidal field of the Galaxy and encounters with binary stars. This would imply that clusters which are close to disruption have a core rich in compact and giant stars (Takahashi \& Portegies Zwart 2000).

Recently, evidence of an open cluster remnant has been discussed by Bassino et al. (2000). They studied the relatively high latitude concentration of stars M73 (NGC6994) and derived an age of 2-3 Gyr from CMDs. They found a significant number density contrast with respect to the galactic field  CMD predicted by count models in the area.  Carraro  (2000) does not favour the object as an open cluster or as a remnant. 
At any rate, if M73 is a physical system the open cluster classification is certainly not adequate. Let us then suggest the acronym POCR - Possible Open Cluster Remnant. As an ongoing study of this neglected class of interesting objects and in view of future  CMDs to determine parameters such as reddening, age and distance, we present a list of  candidates. We  discuss their possible physical nature by checking whether  they present a significant number density contrast with respect to their fields. In Section 2 we present the sample. In Section 3  we analyze the significance of the excesses of  stars by means of (i) equal solid angle counts in the object and field areas, and 
(ii) galactic model counts.
In Section 4 we discuss the results. Finally, in Section 5 we present the concluding remarks.

\section{The sample}

In trying to identify fossil cluster remains we browsed  relatively high galactic latitude ($|b| >$ 15$^{\circ}$) zones by means of
1st and 2nd generation Digitized Sky Survey Images and Guide Star Catalog maps,  searching for poorly populated objects  described as star clusters in open cluster catalogues (Alter at al. 1970, Lyng\aa  \,1987, Lauberts 1982), and also neglected NGC or IC entries originally described as clusters. Although cluster remnants must be common at low galactic latitudes we chose $|b| >$ 15$^{\circ}$ because of lower field contamination, also avoiding young disk and dust distribution details which affect models of low latitude zones (Section 3).

Table 1 lists the 34 POCRs and two comparison clusters, by columns: (1) object designation, (2) and (3) galactic coordinates, (4) and (5) J2000 equatorial coordinates, (6) and (7) the major and minor diameters, and finally (8) E(B-V)$_{FIR}$ reddening values. We measured coordinates of the POCR centers and diameters on Digitized Sky Survey images.  The list is not intended to be extensive or complete, reflecting only the most interesting objects found in the present search. We obtained the E(B-V)$_{FIR}$ reddening values from  Schlegel et al. (1998) dust emission reddening maps. From the study of star cluster directions Dutra \& Bica (2000) called attention to the fact that  Schlegel et al.'s reddening values reflect foreground and background dust contributions. Most objects in Table 1 have relatively high galactic latitudes and their reddening values are quite low, probably corresponding  to foreground dust in the pathsight. The only exception is NGC1901, which is projected onto the LMC disk and the high reddening value in Table 1 arises mainly from LMC dust (Dutra \& Bica 2000).

 Some POCRs present differences between their major and minor diameters (Table 1) indicating ellipticity. Numerical simulations by Terlevich (1987) and  Portegies Zwart et al. (2000) have shown that stars escape from the cluster due to the Galaxy tidal field preferentially through the lagrangian points located on the vector connecting the cluster to the Galactic centre. This effect tends to flatten the clusters in the direction perpendicular to the Galactic Plane. This might explain some cases of ellipticity, considering also  projection effects. Ellipticity can also be explained by the long-term evolution of binary star clusters (de Oliveira et al. 2000).

In the following we comment on  some objects. The designation NGC 1963  is often applied to an edge-on galaxy $\approx$ 11$^{\prime}$ east of the POCR. However the New General Catalogue  describes a star cluster, and the original position (available e.g. in  Sulentic \& Tifft  1973) coincides with that of the POCR. Lauberts (1982) correctly identified the present object (NGC 1963 or ESO363SC5) and the galaxy (IC 2135 or ESO363G7). On the other hand the POCR NGC 2314A has a neighbouring bright galaxy (NGC 2314 itself) located $\approx$ 5$^{\prime}$ north. The New General Catalogue description clearly refers to the galaxy, but the designation NGC 2314 has been applied to the stellar concentration (Alter et al. 1970), which is also known as OCl-374. We suggest the extension 'A' to the POCR designation. 
NGC 2664, NGC 1557 and NGC 1641 are  poorly populated concentrations of stars reported as  clusters
in the NGC catalogue, but not included in modern open cluster catalogues. NGC 2664 is projected just a few cluster diameters  away from the populous intermediate age open cluster M67, an old disk template spatially located
quite high in the plane (z = 430 pc - see Mermilliod's (1996) WEBDA Open Cluster database for essential cluster parameters in the Web interface
 {\it http://obswww.unige.ch/webda}).  
 NGC 1901 is an open cluster or stellar group (Sanduleak \& Philip 1968) that is projected to the north of the LMC bar, whereas  NGC 1641 and NGC 1557 are projected onto the outer parts of the LMC. If physical stellar systems, these objects are clearly related to the Milky Way due to the brightness of their stars.  The close projection of some open clusters and POCRs also leaves the possibility of future investigations, such as common origin, early cluster pairs or multiplets, fossil associations, or perhaps debris of  dwarf galaxy captures.

\begin{table*}
\caption[]{The sample}
%\label{tab1}
\renewcommand{\tabcolsep}{0.5mm}
\renewcommand{\arraystretch}{0.8}
\begin{tabular}{lccccccccccccc}
\hline\hline
Name&$\ell$&{\it b}& RA(2000) & Dec(2000)&D & d &E(B-V)$_{FIR}$&B$_{lim}$&Sol.Ang.&Object&Model&GSC&GSC\\
&($^{\circ}$)&($^{\circ}$)&h:m:ss~~ &$^{\circ}$:$^{\prime}$~~:$^{\prime\prime}~$&($^{\prime}$)&($^{\prime}$)~~&&&(arcmin$^2$)&counts&counts&counts&std\\
\hline
ESO464SC9&15.92 &-39.43 &20:59:37& -29:23:12  &  4  &  3 &0.10&        16.5 &        12.0 &      3$^*$(13) &        2.48 &       0.93$^*$ &       1.17$^*$ \\
M73,NGC6994,Cr426,OCl-89&35.71 &-33.93 &20:58:55& -12:38:03 &   9  &  9 &0.05&        12.5 &        81.0 &           5 &        1.30 &        1.01 &        1.11\\ 
NGC6863&38.27 &-17.99 &20:05:07&  -3:33:20&    2 & 1.5&0.29&        15.5 &         3.0 &           3 &        1.00 &        0.37 &        0.70 \\
DoDz6,OCl-129&61.58 & 40.36 &16:45:24 & 38:21:00   &  6 & 3.5&0.02&        13.0 &        21.0 &           5 &        0.39 &        0.50 &        0.66 \\
NGC7036&64.54 &-21.44 &21:10:02 & 15:31:05  &  4 &   3 &0.08&        14.5 &        12.0 &           5 &        1.48 &        1.75 &        1.37\\
NGC7772,OCl-230&102.73 &-44.27 &23:51:46&  16:14:49&    3 &   2&0.04&        13.5 &         6.0 &           4 &        0.13 &        0.13 &        0.34 \\
NGC5385&118.19  &40.38 &13:52:27  &76:10:27  &  8  &  6 &0.04&        12.0 &        48.0 &           7 &        0.39 &        0.54 &        0.76 \\
Cr21,OCl-371&138.73 &-33.99 & 1:50:11 & 27:04:00&    9&  7.5&0.07&        11.0 &        67.5 &           7 &        0.26 &        0.34 &        0.57 \\
NGC2314A,OCl-374&139.58 & 27.35 & 7:10:12 & 75:12:15 & 4.5 & 3.5 &0.04 &        15.5 &       15.75 &           6 &        2.02 &        1.57 &        1.50\\
NGC2408&143.62 & 29.46 & 7:40:09 & 71:39:20 &  24 &  20 &0.02&        11.0 &       480.0 &          13 &        2.43 &        2.61 &        1.59 \\
DoDz1,OCl-287&158.61 &-37.43 & 2:47:30 & 17:16:00 &  14  &  9 &0.10&        11.0 &       126.0 &           5 &        0.48 &        0.56 &        0.82 \\
NGC1498&203.62 &-43.32 & 4:00:18& -12:00:55&  2.2&  1.8&0.05&        14.0 &         4.0 &           2 &        0.11 &        0.01 &        0.10 \\
NGC2664&214.34  &31.31  &8:47:11 & 12:36:10 &   8  &  8 &0.02&        11.3 &        64.0 &           3 &        0.39 &        0.53 &        0.70 \\
NGC2017,ESO554**22&221.62 &-23.70  &5:39:17 &-17:50:50&    6 &   5 &0.06&        12.0 &        30.0 &           5 &        0.40 &        0.25 &        0.48 \\
ESO486SC45&226.07 &-30.80  &5:16:43 &-24:02:17 & 3.5 & 3.5 &0.03&        15.5 &       12.25 &           8 &        1.35 &        1.44 &        1.17 \\
ESO489SC1&232.92 &-21.41  &6:04:58 &-26:44:05 &  11 &  11 &0.03&        13.2 &       121.0 &          12 &        4.89 &        4.02 &        1.90 \\
ESO425SC6&235.39 &-22.28  &6:04:50 &-29:10:59 &   6 &   6 &0.03&        15.0 &        36.0 &          13 &        4.67 &        4.74 &        2.23 \\
ESO425SC15&236.37 &-20.35  &6:14:35 &-29:22:30 &   6 &   6 &0.04&        14.5 &        36.0 &          10 &        3.80 &        3.84 &        1.89 \\
ESO424SC25&237.72 &-26.37  &5:49:49 &-32:28:20 &   9 &   7 &0.04&        13.0 &        63.0 &           9 &        1.71 &        1.59 &        1.33 \\
ESO426SC26&239.62 &-16.51  &6:36:18 &-30:51:30 &   7 &   7 &0.11&        14.5 &        49.0 &          15 &        6.47 &        5.85 &        2.52 \\
NGC1891,ESO362?20&239.69 &-32.87  &5:21:25 &-35:44:29 &  15 &  10&0.04&        12.5 &       150.0 &           6 &        2.19 &        1.43 &        1.20 \\
NGC1963,ESO363SC5&240.99 &-30.86  &5:32:17 &-36:23:30 &  13 &  13 &0.03&        13.0 &       169.0 &          13 &        3.81 &        3.06 &        1.90 \\
ESO437SC61&273.06 & 26.22 &10:48:03 &-29:23:28 &   5  &  4 &0.07&        16.0 &        20.0 &       5$^*$(9) &        4.35 &        2.20$^*$ &        1.38$^*$ \\
ESO245SC9&273.76 &-67.48  &1:53:43 &-45:57:16 &  14  & 12 &0.02&        13.5 &       168.0 &          11 &        2.41 &        2.36 &        1.50 \\
ESO570SC12&274.06 & 35.92 &11:12:12 &-21:19:13 &  13  & 13 &0.03&        12.8 &       169.0 &          10 &        3.01 &        2.15 &        1.40 \\
NGC1252,ESO116?11&274.08 &-50.83  &3:10:49 &-57:46:00 &  14 & 11&0.02&        12.0 &       154.0 &           8 &        1.02 &        1.04 &        1.00 \\
ESO502SC19&276.02 & 30.68 &11:08:07 &-26:43:51  &  3 &   3 &0.05&        15.0 &         9.0 &           3 &        0.87 &        0.63 &        0.75 \\
NGC1641,ESO84SC24&277.20 &-38.31  &4:35:32 &-65:45:03  & 11 &   9 &0.04&        13.5 &        99.0 &          11 &        2.64 &        2.86 &        1.72 \\
NGC2348,ESO88SC1&278.14 &-23.80  &7:03:03 &-67:24:44 &  18 &  10 &0.12&        13.0 &       100.0 &           5 &        3.28 &        2.57 &        1.61 \\
NGC1557,ESO55**15&283.77 &-38.26  &4:13:11 &-70:28:18  & 21 &  17 &0.10&        12.0 &       357.0 &           8 &        3.19 &        2.69 &        1.57 \\
ESO442SC4&298.40  &33.30 &12:34:05 &-29:24:38  & 11 &  11 &0.08&   12.0-14.0 &       121.0 &          11 &        4.32 &        2.98 &        1.92 \\
IC1023,ESO385SC39&324.95 & 22.71 &14:32:25 &-35:48:13 &   5 &   5 &0.07&        15.0 &        25.0 &          13 &        4.72 &        3.19 &        1.55 \\
ESO141SC47&338.95 &-26.17 &19:18:03 &-57:54:16 &   7 &   5 &0.06&        14.5 &        35.0 &          11 &        3.93 &        2.62 &        1.59 \\
ESO282SC26&355.01 &-21.89 &19:13:52 &-42:38:58  & 15 &  13 &0.08&        12.5 &       195.0 &          12 &        6.26 &        3.43 &        1.70 \\
\hline\hline
&&&Comparison &Clusters&&&\\
\hline
NGC3680,Mel-106&286.76  &16.92 &11:25:38 &-43:14:30 &  15  & 15 &0.09&   10.9-12.5 &       225.0 &          20 &        6.14 &        3.55 &        2.14 \\
NGC1901,Bok 1,OCl-791.1&279.03 &-33.60  &5:18:11 &-68:27:00 &  19 &  19 &0.36&        10.5 &       361.0 &           9 &        0.91 &        2.63 &        1.88\\
\hline
\end{tabular}
\begin{list}{}
\item *  GSC not deep enough to reach assumed B$_{lim}$. Number in parenthesis comes from DSS visual counts.
\end{list}
\end{table*}

We show in Figure 1 the angular distribution of the 34 sample objects and the two comparison clusters. POCRs can be basically  found in any direction
along the plane. The number of POCRs is small but concentrations could be present in the third Quadrant, which would suggest that several objects therein had common origin. We emphasize that our survey of POCRs was based on previously catalogued objects, and it would be important to carry out a systematic search for new objects with similar characteristics in the same part of sky, as well as for $|b| <$ 15$^{\circ}$.

\begin{figure} 
\resizebox{\hsize}{!}{\includegraphics{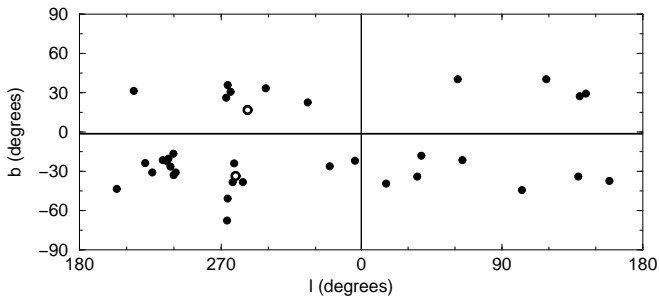}}
\caption[]{Angular distribution of the 34 POCRs (filled circles) together with the comparison clusters NGC\,3680 and NGC\,1901 (open circles). Galactic Plane and Minor Axis are indicated by solid lines.}
\label{fig1}
\end{figure}

\begin{figure} 
\resizebox{\hsize}{!}{\includegraphics{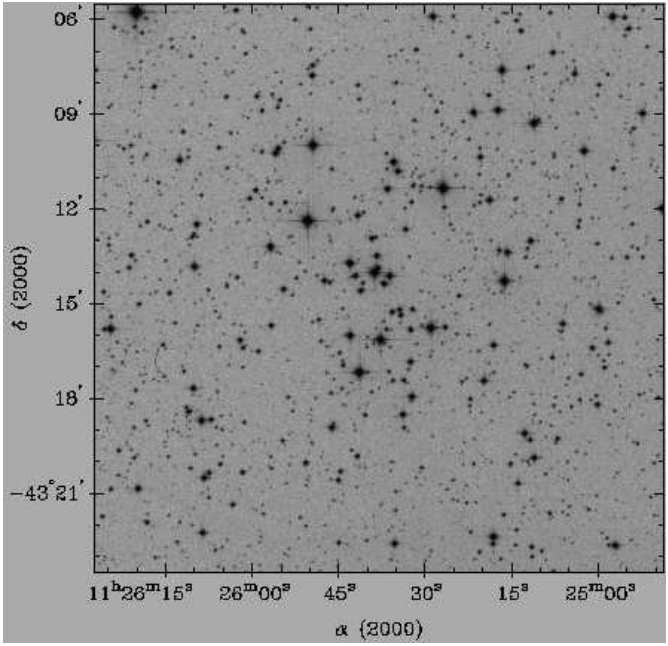}}
\caption[]{18$\times$18 arcmin$^2$ XDSS image of the depleted low Main Sequence open cluster NGC3680}
\label{fig1}
\end{figure}

\begin{figure} 
\resizebox{\hsize}{!}{\includegraphics{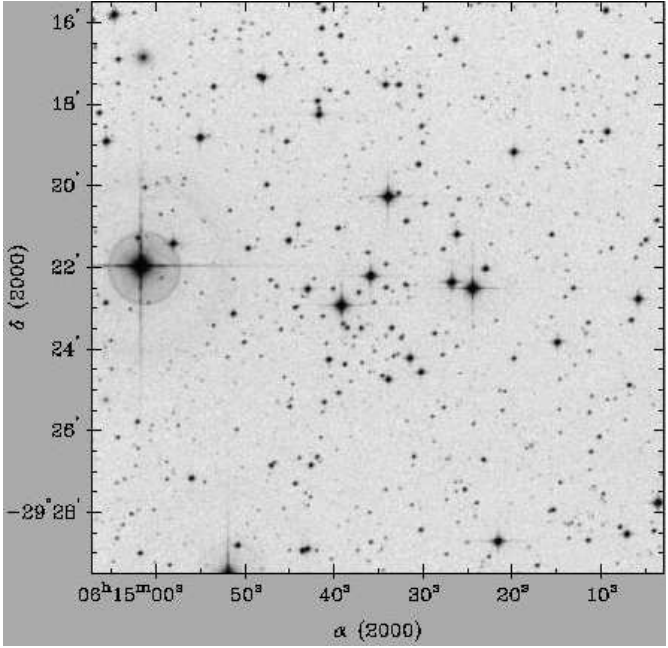}}
\caption[]{14$\times$14 arcmin$^2$ XDSS image of the POCR ESO425SC15}
\label{fig1}
\end{figure}

\begin{figure} 
\resizebox{\hsize}{!}{\includegraphics{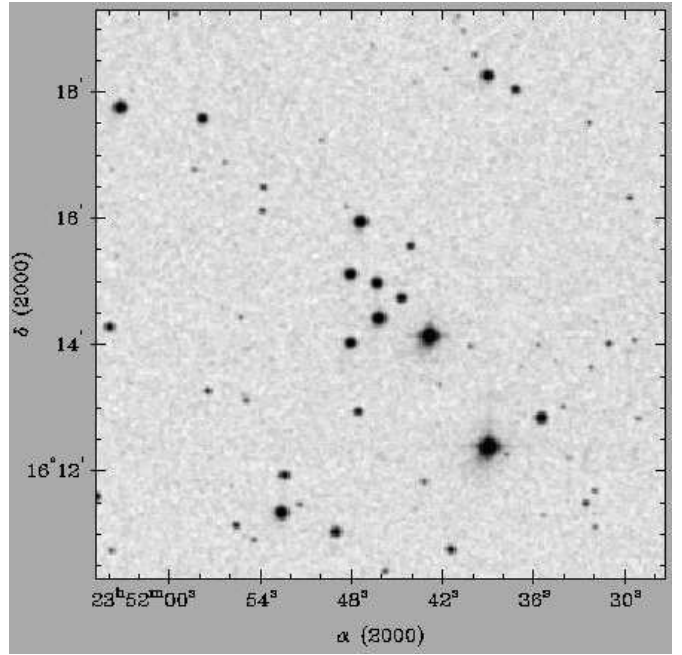}}
\caption[]{9$\times$9 arcmin$^2$ DSS image of the POCR NGC7772}
\label{fig1}
\end{figure}

\begin{figure} 
\resizebox{\hsize}{!}{\includegraphics{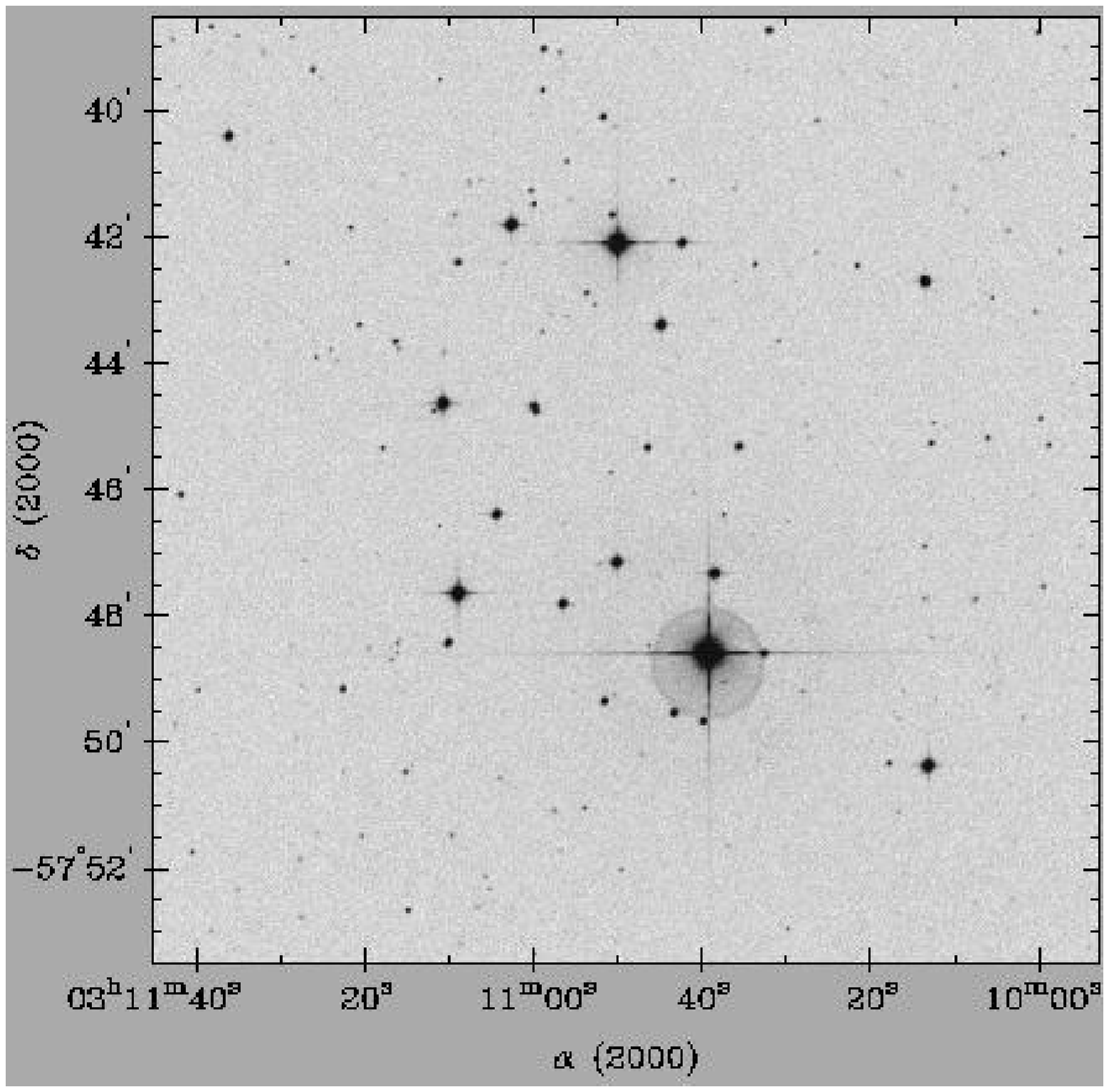}}
\caption[]{15$\times$15 arcmin$^2$ XDSS image of the POCR NGC1252}
\label{fig1}
\end{figure}

Figure 2 shows a 2nd generation Digitized Sky Survey (XDSS) image of the comparison cluster NGC 3680 which has low MS depletion (Anthony-Twarog et al. 1991). The WEBDA database provides a distance of  735 pc which combined to the angular dimensions in Table 1 imply a linear diameter of 3.2 pc. The other comparison cluster NGC 1901 (Table 1) has an age of 830 Myr according to WEBDA. The distance is 415 pc implying a linear diameter of 2.3 pc. NGC 3680 and NGC 1901 are  probably evolving into POCRs. Figure 3 shows a XDSS image of ESO425SC15 which resembles NGC 3680, but seems to be more depleted in terms of faint stars. Figure 4 shows a 1st generation image (DSS) of  NGC 7772 which is a more compact POCR. Finally, Figure 5 shows a XDSS image of NGC 1252 which is a POCR with a rather loose morphology. The New General Catalogue describes  NGC 1252 as a star cluster with 18-20 stars. This description is compatible with the only concentration of stars in the area, which is the present POCR with dimensions 14$^{\prime} \times$ 11$^{\prime}$ (Table 1). Bouchet $\&$
 Th\'e (1983) carried out photometry in a region with diameter
 $\approx$ 1$^{\circ}$ centred on the bright carbon star TW Horologii. The POCR
 is located at the edge of Bouchet $\&$ Th\'e's large region. The only star in the POCR area considered to be member of the cluster as interpreted by Bouchet $\&$ Th\'e is BT1. Recently Baumgardt (1998)
 discussed 12  bright stars in Bouchet $\&$ Th\'e's large region based on
 the ACT and Hipparcos catalogues and concluded that those stars do not make
 up a cluster. The present POCR which is to be identified with NGC 1252 has not yet been explored.

\section{Number density contrast analysis}

Guide Star Catalog 1.1 (hereafter, GSC) maps and/or  DSS/XDSS  images clearly show that the present
objects stand out from their surrounding fields. In the following we assess the statistical significance of the excess projected number density of stars represented by the POCRs that allowed previous authors to identify them on photographic plates. We use two independent methods to analyze the number density contrast: (i) comparing  star counts  from DSS images in the cluster region up to a limiting magnitude at which the  cluster prevails over the  neighbouring field, with  theoretical star counts based on a Galactic Structure Model predicted up to the  same magnitude; (ii) the same cluster counts  are compared to counts in  neighbouring GSC fields with solid angle equal to the cluster's. The magnitude range responsible for the excess stars was visually defined for each  POCR using DSS and XDSS images and the GSC positions and magnitudes.

 The adopted Galactic Model has been previously described in more detail
(Santiago et al. 1996, Reid \& Majewski 1993). It includes 3 structural
components: a thin disk, a thick disc and a spheroid. Given the
typical magnitude limit for the cluster stars (V$ \approx $15), 
our relevant star counts are almost entirely dominated by the thin disk. Its stellar density profile is described by a double exponential, one along the
plane of the disk and  the other perpendicular to it. As for the luminosity
function, it is very well constrained 
for disc stars with $M_{V} \approx $ 10, which 
will largely dominate the field star number counts in the DSS images and GSC
catalogue (Wielen et al. 1983, Bessel \& Stringfellow 1993).
We therefore anticipate no model uncertainties propagating into the predicted
number counts of field stars.

The GSC fields were extracted using the task REGIONS in the STSDAS package inside {\bf IRAF}.
With the positions and magnitudes of all GSC stars within $12~\theta_{clus}$,
where $\theta_{clus}$ is the apparent major diameter of each
POCR, defined from inspection
of the DSS images, we randomly selected 100 centers within 
this region and counted
the number of GSC stars found within the same solid angle and magnitude range
defined for the  POCR. The procedure is illustrated  by Figure 6. The model and GSC counts were taken from a solid angle $\omega = D~d$,
which corresponds to a rectangle that encompasses the cluster ellipse,
therefore being larger than that of
the cluster by a factor $4 / \pi$. This conservative procedure of
estimating expected field star counts is an attempt to compensate for
uncertanties in the defined cluster boundaries.

\begin{figure} 
\resizebox{\hsize}{!}{\includegraphics{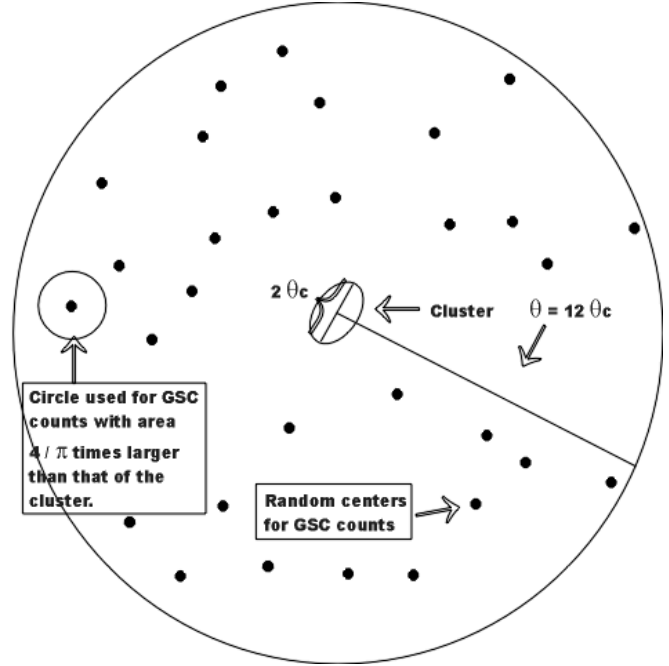}}
\caption[]{Schematic representation of the procedure employed to count stars in the object and surrounding field samplings.}
\label{fig1}
\end{figure}

The results of both experiments are given in  columns (9) through (14) of Table 1. We show the limiting magnitude (in two cases bright field stars also occurred so that a bright limit is also shown), solid angle 
and number of stars found for the POCR. The next column lists the expected number
of stars from the model in the same region.
Finally, we list the average and dispersion of the GSC star counts 
in randomly placed areas in the cluster's neighbourhood.

Most entries clearly show an excess of observed stars relative to both
model expected values and GSC counts. This is clearly seen in the upper panel of Figure 7, where cluster counts are plotted against the average GSC counts. Exceptions are ESO464SC9, ESO437SC61
and NGC2348, but in the former two the GSC is not deep enough to reach the assumed B$_{lim}$ (Table 1). For all other  POCRs, the observed counts well exceed
the dispersion of GSC counts relative to the mean value. The high concentration of stars in M73 (NGC 6994) occurs in a 2$^{\prime}$ $\times$ 2$^{\prime}$ region (Bassino et al. 2000, Carraro 2000). However the diameter in Table 1 of  9$^{\prime}$ $\times$ 9$^{\prime}$ corresponds to a halo region with probable members (Bassino et al. 2000). This larger area provides a conservative estimate of the number density contrast, which is significant (Table 1). The distance of 620 pc (Bassino et al. 2000)  implies linear diameters of 1.2 pc and 0.4 pc for the halo and core regions respectively. These estimates can be compared with those of the dynamically evolved clusters NGC 3680 and NGC 1901 (Section 2), resulting smaller for the POCR.

The two estimates of background counts, model
and GSC, agree quite closely, as shown in the lower panel of Figure 7. The reason why model counts lie, in average, slightly above the GSC ones is likely
caused by incompleteness effects affecting the latter. This consistency 
in the two independent estimated numbers of field stars and the high number density contrasts support the possibility of the  POCRs being real physical systems.

\begin{figure} 
\resizebox{\hsize}{!}{\includegraphics{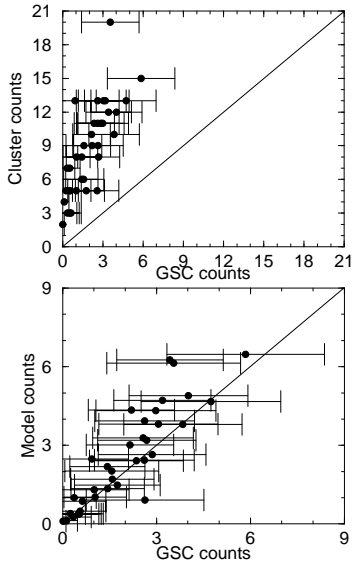}}
\caption[]{Count comparisons with those from the Guide Star Catalogue. Upper panel: counts in the object solid angle. Lower panel: Galactic Model counts. Error bars indicate 1$\sigma$ deviations from average GSC counts.}
\label{fig1}
\end{figure}

\section{Discussion}

The number density contrast of stars in the objects  with respect to the field  resulted significant (Section 3), revealing a real concentration, and this is the first step towards establishing a possible physical nature. The construction of the spatial distribution will require CMDs, but their high latitude locations give some hints on their origin as galactic subsystems. Note that M73's age and position (Bassino et al. 2000) place it in the old disk. The available similar properties and angular distributions suggest the old disk as the likely origin for most POCRs in Table 1. It is not unexpected from the high latitudes that a globular cluster remnant be present. Spectroscopic and velocity studies might readily reveal halo characteristics. It is not excluded either that the sample includes a thick disk (Gilmore \& Reid 1983, Yoshii et al. 1987) cluster remnant.
The scale height attributed to the thick disk is  about 1 kpc (Kerber et al. 2000), and it should be very old, near the upper age limit of the old disk. 
In this scenario the thick disk would be a direct halo/disk transition as gas settled into the disk forming stars. Alternatively, the thick disk could be a thin  disk population heated by some internal or external mechanism such as a bar  or by dwarf galaxy captures throughout  the Galaxy history (Freeman 1992). In the latter scenario clusters with t \gtsima 500 Myr would be expected at relatively large heights.  Some orbital debris might be identified by means of high latitude clumps of star clusters and/or their fossils, which also increases the importance of Table 1 objects.

The evolutionary stage and duration of the POCR phase in the lifetime of a star cluster can be inferred from available numerical models. Assuming that a cluster becomes significantly depopulated with a POCR appearance by losing 2/3 of its stars, simulations of clusters of initial mass 1000 M$_{\odot}$ with various mass spectra and concentration degrees (Terlevich 1987) show that the depopulated phase lasts $\Delta$t$\approx$ 200 Myr for a total lifetime of $\approx$ 1 Gyr. The timescale from the depopulated phase until  dissolution would be about 400 Myr. Portegies Zwart et al. (2000) computed models for 1500 M$_{\odot}$ considering a range of galactocentric distances. For a distance R$_{GC}$ $\approx$ 12 kpc the cluster is more stable and the POCR phase could last much longer than above. For R$_{GC}$ $\approx$ 6 kpc  the timescales are comparable to those of the 1000 M$_{\odot}$ clusters computed by Terlevich (1987). For less massive model clusters (Terlevich 1987) the POCR phase would be typically 1/3 to 1/4 of the cluster lifetime. Considering a catalogued population of open clusters of $\approx$ 1000 (Lyng\aa\, 1987, Alter et al. 1970) as many as 500 POCRs would be expected in all galactic latitudes. Considering the known sample of intermediate age open clusters (Dutra \& Bica 2000) we expect $\approx$ 50 POCRs of such ages. Clearly, a complete census of POCRs requires additional surveys, in particular for lower galactic latitudes. Detailed observations of each object are necessary to assess their physical properties and as a consequence to shed more light on the last stages of star cluster dynamical evolution.

\section{Concluding remarks}

We presented a list of 34 neglected entries from star cluster catalogues located at relatively high galactic latitudes ($|b| >$ 15$^{\circ}$)  which may be late stages of cluster dynamical evolution.  Although underpopulated with respect to usual open clusters, we showed that they still present a high number density contrast with respect to the galactic field, as verified by means  of (i) predicted model counts from different galactic subsystems in the same direction, and (ii) Guide Star Catalog equal solid angle counts for the object and surrounding fields.
The sample will be useful for followup studies, aimed at  verifying their physical nature. Photometry and spectroscopy are required to determine fundamental parameters such as reddening, distance, age, radial velocities, membership and chemical abundances. Some of these objects might be clusters in the process of disruption or  their fossil remnants. The dynamical state  of the physical objects in the present sample may be inferred from comparisons of the cluster remnant and surrounding field luminosity functions, searching  for depletion effects.

An important population of possible open cluster remnants is likely to  exist. They may survive significant amounts of time as depopulated systems before dissolution. Simple arguments based on available numerical models and catalogued open clusters suggest that several hundreds can be expected. Systematic surveys to find new candidates and numerical models to explore in detail the evolved dynamical stages are encouraged for a better understanding of this so far overlooked object class.

\begin{acknowledgements}

We use data from The Guide Star Catalog 1.1 and Digitized Sky Survey which were produced at the Space Telescope Science Institute under U.S. Government grants NAS5-26555 and NAG W-2166, respectively. These data are based on photographic data obtained using the Oschin Schmidt Telescope on Palomar Mountain and the UK
Schmidt Telescope.
We thank the referee  Dr. Simon Portegies Zwart, for interesting remarks.
We acknowledge support from the Brazilian institution CNPq.
 
\end{acknowledgements}


\begin{thebibliography}{}
\bibitem[]{} Ahumada A.V., Clari\'a J.J., Bica E., Piatti A.E. 2000, A\&AS, 141, 79
\bibitem[]{} Alter G., Ruprecht J., Va\'nysek J. 1970, Catalogue of Star Clusters and Associations, eds. G. Alter, B. Bal\'azs, and J. Ruprecht (Akademiai Kiado, Budapest)
\bibitem[]{} Anthony-Twarog B.J., Heim E.A., Twarog B.A., Caldwell N. 1991, AJ, 102, 1056
\bibitem[]{} Bassino L.P., Waldhausen S., Mart\'inez R.E. 2000, A\&A, 355, 138
\bibitem[]{} Baumgardt H. 1998, A\&A, 340, 402
\bibitem[]{} Bessel M.S., Stringfellow G.S. 1993, ARA\&A, 31, 433
\bibitem[]{} Bica E., Ortolani S., Barbuy B. 1999, A\&AS, 136, 363
\bibitem[]{} Bouchet R.,Th\'e P.S. 1983, PASP, 95, 474
\bibitem[]{} Carraro G. 2000 A\&A, 357,145
\bibitem[]{} de la Fuente Marcos R. 1997, A\&A, 322, 764
\bibitem[]{} de Oliveira M.R., Bica E., Dottori H. 2000, MNRAS, 311, 589
\bibitem[]{} Dutra C.M., Bica E. 2000, A\&A, 359, 347
\bibitem[]{} Freeman K.C. 1992, The Stellar Population of Galaxies, ed. B. Barbuy, A. Renzini, (Kluwer:Dordrecht), p. 65
\bibitem[]{} Friel E.D. 1995 ARA\&A, 33, 381
\bibitem[]{} Gilmore G., Reid N. 1983, MNRAS, 202, 1025
\bibitem[]{} Kerber L.O., Javiel S., Santiago B.X. 2000, A\&A, submitted
\bibitem[]{} Lauberts A. 1982 The ESO/Uppsala Survey of the ESO (B) Atlas, European Southern Observatory, Garching bei M\"unchen 
\bibitem[]{} Lyng\aa\,  G. 1987, Catalog of Open Star Cluster Data (Strasbourg:CDS)
\bibitem[]{} McClure R.D., Hesser J.E., Stetson P.B., Stryker L.L. 1985, PASP, 97, 665
\bibitem[]{} Mermilliod, J.C. 1996, In: The origins, evolution, and destinies of binary stars in clusters. ASP Conf. Ser. 90, p. 475
\bibitem[]{} Ortolani S., Barbuy B., Bica E. 1999, A\&AS, 136, 237
\bibitem[]{} Portegies Zwart S.F., McMillan S.L.W., Hut P., Makino J. 2000, MNRAS, submitted (astro-ph/0005248)
\bibitem[]{} Reid I.N., Majewski S.R. 1993, ApJ, 409, 635
\bibitem[]{} Sanduleak N., Philip A.G.D. 1968, AJ, 73, 566 
\bibitem[]{} Santiago B.X., Gilmore G., Elson R.  1996, MNRAS, 281, 871
\bibitem[]{} Schlegel D.J., Finkbeiner D.P., Davis M. 1998, ApJ, 500, 525
\bibitem[]{} Sulentic J.W., Tiff W.G. 1973, The Revised New General Catalogue of nonstellar astronomical objects,(The University of Arizona Press, Tucson)
\bibitem[]{} Terlevich E. 1987, MNRAS, 224, 193
\bibitem[]{} Takahashi K., Portegies Zwart S.F. 2000, ApJ, 535, 759
\bibitem[]{} Wielen R. 1971, A\&A, 13, 309
\bibitem[]{} Wielen R., Jahreiss H., Kr\"uger R. 1983, in IAU Coll.76, {\it Nearby Stars and the Stellar Luminosity Function}, eds A.G.D. Philip \& A.R. Upgren, (L. Davis Press: Schenectady), p. 163  
\bibitem[]{} Yoshii Y., Ishida K., Stobie R.S. 1987, AJ, 93 323

\end{thebibliography}
\end{document}